\pdfoutput=1

\documentclass[%
 aip,
 amsmath,amssymb,
 reprint,%
]{revtex4-1}

\usepackage{graphicx}
\usepackage{dcolumn}
\usepackage{bm}
\usepackage[normalem]{ulem}
\usepackage{xcolor}

\usepackage[utf8]{inputenc}
\usepackage[T1]{fontenc}
\usepackage{mathptmx}
\usepackage{etoolbox}

\makeatletter
\def\@email#1#2{%
 \endgroup
 \patchcmd{\titleblock@produce}
  {\frontmatter@RRAPformat}
  {\frontmatter@RRAPformat{\produce@RRAP{*#1\href{mailto:#2}{#2}}}\frontmatter@RRAPformat}
  {}{}
}%
\makeatother
\begin{document}

\preprint{AIP/123-QED}

\title[Modeling creeping flows in porous media using regularized Stokeslets]{Modeling creeping flows in porous media using regularized Stokeslets}
\author{Suraj Kumar Kamarapu}
\affiliation{ 
	Department of Mechanical Engineering, University of Utah, Salt Lake City, UT 84112 USA.
}
\author{Mehdi Jabbarzadeh}%
\affiliation{ 
	Department of Mechanical Engineering, University of Utah, Salt Lake City, UT 84112 USA.
}
\author{Henry Chien Fu}
 \email{henry.fu@utah.edu}
\affiliation{ 
Department of Mechanical Engineering, University of Utah, Salt Lake City, UT 84112 USA.
}%
 \homepage{http://hfu.mech.utah.edu}

\date{\today}

\begin{abstract}
Flows in porous media in the low Reynolds number regime are often modeled by the Brinkman equations. Analytical solutions to these equations are limited to standard geometries. Finite volume or element schemes can be used in more complicated geometries, but become cumbersome when there are moving boundaries that require frequent remeshing of the domain. In Newtonian fluids, the method of regularized Stokeselets has gained popularity due to its ease of implementation, including for moving boundaries, especially for swimming and pumping problems.  While the corresponding method of regularized Brinkmanlets can be used in a domain consisting entirely of Brinkman medium, many applications would benefit from an easily implemented representation of flow in a domain with heterogeneous regions of Brinkman medium and Newtonian fluid. In this paper, we model flows in porous media by scattering many static regularized Stokeslets randomly in three dimensions to emulate the forces exerted by the rigid porous structure. We perform numerical experiments to deduce the correspondence between the chosen density and blob size of regularized Stokeslets in our model, and a Brinkman medium. 
\end{abstract}

\maketitle


\section{\label{sec:intro}Introduction\protect}
Flow through porous media is encountered in many natural and industrial systems \cite{Xue2020}. A quantitative description of flows in porous media is extremely important for understanding flow phenomena or designing such systems. Mathematical modeling of such flows has been an active area of research in many fields, including, but not limited to, medicine \cite{Vafai2010}, fluid mechanics \cite{Philip1970}, hydrology \cite{Carrillo2019}, geophysics \cite{Riaz2006}, and soil mechanics \cite{Sheldon2006}. A volume-averaged description of flow parameters and global description of porous structure (\textit{e.g.,} porosity, permeability) helps avoid the characterization of spatially intricate flows through small-scale porous structures. Attempts to derive equations describing flows through porous media date back to 1856 when Darcy \cite{Darcy1856} reported empirical relations based on experimental observations relating the averaged fluid velocity to the pressure drop via a linear relationship. This model presumes that the averaged flow field is uniform in the domain and the drag offered by the porous structure dominates viscous shear forces in the fluid. It fails to model flows in a highly porous domain and near boundaries \cite{Neale1974}. For non-uniform flows through porous media, the Brinkman equation was proposed in 1949 \cite{Brinkman1949} to describe flows past a dense swarm of static spherical particles.  
Here, the viscous forces due to velocity gradients in the fluid flow become comparable to the drag force exerted by the porous structure leading to the Brinkman equations \cite{Brinkman1949}: 
\begin{eqnarray}
-\nabla p &=& -\mu\nabla^2\textbf{u} + \mu\alpha^2\textbf{u} ,\,\,\,\,\, \nonumber \\
\nabla \cdot \textbf{u} &=& 0 ,\label{brinkman}
\end{eqnarray}
where $p$ is the averaged pressure,  $\textbf{u} $ is the averaged fluid velocity, $\phi$ is the porosity (volume fraction of fluid), $\mu$ is the dynamic viscosity of the interstitial fluid, and $\alpha$ is the resistance of the porous medium ($\alpha^{-2}$ is the permeability). Various theoretical studies have shown that this model accurately describes the flow in dilute porous media \cite{Howells1974, Neale1974, Rubinstein1986, Durlofsky1987, Liu2005}. The value of $\alpha$ sets the length scale $(\alpha^{-1})$, called Brinkman screening length, over which the fluid flow decays in a given porous medium. In the absence of porous structure in the fluid flow, $\alpha\rightarrow0$ and the Brinkman equation reduces to Stokes equation $\left(-\nabla p + \mu \nabla^2 \textbf{u}=0\right)$.   

The Brinkman equation has been used in a variety of applications to mathematically model flows through undeformed gels \cite{Fu2010}, arrays of fixed fibers \cite{Howells1998}, blood clots \cite{Link2020}, and particles in a microfluidic channel \cite{Uspal2014}, as well as how porous media damp flows \cite{Arthurs1998}, and confine active swimmers \cite{Mirbagheri2016,Nganguia2020,Feng1998}. Analytical solutions for the flows obeying Brinkman equations can be derived for problems involving simple geometries such as two-dimensional waving sheets, spheres, or cylinders \cite{Mirbagheri2016,Mahabaleshwar2019,Nganguia2020,Filippov2013}, but they are generally not available for flows involving complicated geometries. 
For domains of arbitrary shapes, numerical solutions can be obtained via finite volume or element schemes, for which the resolution of the discretization dictates the accuracy and computational cost of these numerical solutions. Implementation of these schemes become cumbersome when there are moving boundaries, such as the surface of a swimming bacterium, as it requires frequent remeshing of the domain. If the medium around the moving boundary is a homogeneous Brinkman medium, the method of regularized Brinkmanlets is quite useful \cite{Ho2019, Leiderman2016}, but is invalid when there is a heterogeneity involved. An example of such a heterogeneous mixtures is a porous medium next to a Stokes fluid, as encountered by the bacterium \textit{Helicobacter pylori} swimming in a fluid pocket surrounded by gastric mucus \cite{Mirbagheri2016,Nganguia2020}, or around growing blood clots inside a microfluidic channel \cite{Link2020}. Such heterogeneous domains can be treated using finite volume schemes, but that can become cumbersome as it requires frequent re-meshing of the domain and accounting for complicated boundary conditions along the interface of two fluids. Here we present a computational framework to address fluid flows in such situations using regularized Stokeslets.  

Our basic approach is to model a Brinkman medium by a random arrangement of many static regularized Stokeslets in three dimensions to emulate a rigid (stationary) porous structure. Leshansky \cite{Leshansky2009} implemented a similar approach in which static spherical (or circular in 2D) particles represented a gel-like medium around micro-swimmers. In a  heterogeneous domain, only the portion containing the porous medium is represented by a random spatial arrangement of static regularized Stokeslets. The parameter that defines a porous medium via Brinkman equations is the resistance, $\alpha$. On the other hand, in our proposed model, we define a porous medium using the number density of regularized Stokeslets, $\rho$, and the blob parameter (size of each regularized Stokeslet), $\varepsilon$. To quantify the relationships between the parameters $\alpha$ and $\lbrace\rho, \varepsilon\rbrace$ in describing the same porous medium, we perform numerical experiments using our proposed model and compare to analytical solutions for the Brinkman model.

This paper is organized as follows: In Section II, we review the method of regularized Stokeslets. In Sec. III, we perform numerical experiments involving the proposed model of porous media and match results with the analytical solution obtained using the Brinkman equation. We then connect the input parameters from our model to the resistance of Brinkman medium. In Sec. IV, we discuss the applicability of the proposed porous medium framework to model flows in porous media. 
%
%
\section{Review of regularized Stokeslets}
\label{sec:1}
The method of regularized Stokeslets, introduced in 2001 by Cortez \cite{Cortez2002}, computes the Stokes flow at any point in a domain as the superposition of flow fields generated by forces distributed at material points in a fluid. It uses a Green's function solution, ($\mathbf{u}^\varepsilon, p^\varepsilon$), of Stokes equations in the presence of a force ($\textbf{f}_0$) distributed around position $\textbf{x}_0$ according to the function ${\psi_{\varepsilon}(\textbf{x}-\textbf{x}_0) = \frac{15\varepsilon^4}{8\pi (|\textbf{x}-\textbf{x}_0|^2 + \varepsilon^2)^{7/2}}}$. $\psi_{\varepsilon}$ is a radially symmetric, smooth approximation to a three-dimensional delta distribution with the property that $\int \psi_{\varepsilon}(\textbf{x}) d\textbf{x} = 1$, so that $\psi_\varepsilon(\textbf{x}-\textbf{x}_0)$ is concentrated near $\textbf{x} = \textbf{x}_0$. The approximate size of the distribution is set by the value of the ``blob parameter'' $\varepsilon$. Thus $\mathbf{u}^\varepsilon$  and $p^\varepsilon$ satisfy  
\begin{eqnarray}\label{regstokes}
-\nabla p^\varepsilon + \mu \nabla^2 \textbf{u}^\varepsilon &=& -\textbf{f}_0\psi_ {\varepsilon} (\textbf{x}-\textbf{x}_0) \nonumber \\
 \nabla \cdot \textbf{u}^\varepsilon &=& 0, 
\end{eqnarray}
where $p^\varepsilon$ is the pressure, $\mu$ is the viscosity of the fluid, $\textbf{u}^\varepsilon$ is the fluid velocity, and $\textbf{f}_0$ is the force acting at a point $\textbf{x}_0$. The flow field $\textbf{u}^\varepsilon$ and pressure field $p^\varepsilon$ at a point $\textbf{x}$ are called the regularized Stokeslet and are given by: 
\begin{eqnarray}
	\mathbf{u}^\varepsilon(\mathbf{x}) &=&  \mathbf{S}^{\varepsilon}(\mathbf{x}-\mathbf{x}_0) \cdot \mathbf{f}(\mathbf{x}_0)  \\ 
	p^\varepsilon(\mathbf{x})&=&  \mathbf{P}^{\varepsilon}(\mathbf{x}-\mathbf{x}_0) \cdot \mathbf{f}(\mathbf{x}_0)
\end{eqnarray}
where
\begin{eqnarray}
	\mathbf{S}^{\varepsilon}(\mathbf{r}) &=& \frac{1}{8\pi\mu}\left(\frac{\mathbf{I}(r^2+2\varepsilon^2) + \mathbf{r}\mathbf{r}^T}{(r^2+\varepsilon^2)^{3/2}}\right), \\ 
	\mathbf{P}^{\varepsilon}(\mathbf{r}) &=& \frac{2r^2+5\varepsilon^2}{8\pi(r^2+\varepsilon^2)^{5/2}}\mathbf{r},
\end{eqnarray}
and $\textbf{I}$ is the identity matrix. Crucially, regularizing the point force ($\mathbf{f}_0$) removes any singularity at $\textbf{x}_0$, 
thus simplifying the numerical implementation. The linearity of Stokes equations lets us obtain a solution for multiple forces of the same form, acting at $N$ different locations $\textbf{x}_q$, for $q = 1,2, ..., N$, by superposition of flow fields generated by each of those forces independently. Thus the flow field ($\textbf{u}$) and pressure ($p$) at a point $\textbf{x}$ is given by:
\begin{eqnarray}
\mathbf{u}(\mathbf{x}) &=& \sum_{q=1}^{N} \mathbf{S}^{\varepsilon}(\mathbf{x}-\mathbf{x}_q) \cdot \mathbf{f}(\mathbf{x}_q)  \\ 
p(\mathbf{x}) &=& \sum_{q=1}^{N} \mathbf{P}^{\varepsilon}(\mathbf{x}-\mathbf{x}_q) \cdot \mathbf{f}(\mathbf{x}_q). 
\end{eqnarray} 
Representing the flow external to a body using a surface distribution of regularized Stokeslets is mathematically equivalent to representing the flow via boundary integral equations \cite{Cortez2005}.  Our implementation of the method of Regularized Stokeslets has been previously described \cite{Hyon2012, Martindale2016}.

Due to the ease of its implementation even in intricate three dimensional domains, the method of regularized Stokeslets has become a popular choice to numerically model flows governed by the Stokes equations for swimming microorganisms \cite{Cortez2005,Smith2009,Constantino2016,Constantino2018}, human sperm \cite{Ishimoto2017}, flexible filaments \cite{Olson2013}, microrobots \cite{Gibbs2011, Cheang2014, Meshkati2014a, Fu2015, Ali2017, Cheang2016,Samsami2020, Samsami2020a}, tumor tissues \cite{Rejniak2013}, micropumps \cite{Aboelkassem2013, Aboelkassem2014, Martindale2017}, phoretic flows \cite{Thomas2015}, and plasma membranes \cite{Fogelson2014}. An interconnected lattice network of regularized Stokeslets has been used to develop a computational framework to represent a viscoelastic fluid \cite{Wrobel2014}. 

In this paper, we model porous media by filling the domain occupied by porous media with randomly placed regularized Stokeslets, and connect it with a Brinkman equation description of the same medium.

\section{Numerical experiments} 
The parameter that defines a porous medium via Brinkman equations is the resistance, $\alpha$. On the other hand, in our proposed model, we define a porous medium using the number density of regularized Stokeslets, $\rho$, and the blob parameter (size of each regularized Stokeslet), $\varepsilon$. In order to establish a relation between the two models, we perform two numerical experiments using our proposed model and compare to analytical solutions for the Brinkman model to quantify and validate the relationships between  $\alpha$ and $\{\rho, \varepsilon\}$.
A dimensional analysis suggests that the relationship can be written in the form $\alpha \varepsilon = f( \rho\varepsilon^3, l_1/\varepsilon, l_2/\varepsilon, ...)$, where the $l_i$ are lengthscales defining the geometry of an experiment.

\subsection{Couette flow between two plates}\label{numexp1} 
\vspace{-3mm}
In the first experiment, we consider a steady state flow between two infinitely long plates separated by distance $H$, filled with a dilute porous medium of resistance $\alpha$. The top plate moves to the right with velocity $U$, while the bottom plate is fixed in place.

\begin{figure*}[t]
	\begin{center}
		\includegraphics[width=17cm]{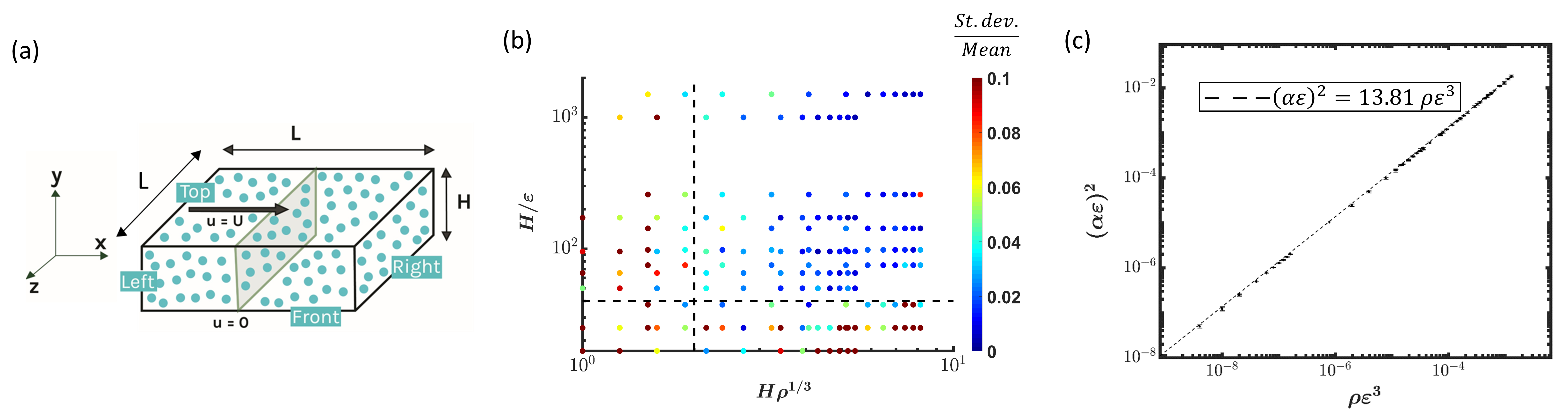}
		\vspace{-12 pt}
		\caption{(\textbf{a}) Box-shaped domain for a portion of the Couette flow driven by the top surface sliding to the right with velocity $U$. (\textbf{b}) Plot showing the ratio of standard deviation to the mean for estimates of $\alpha$ for a given combination of $\{\rho,\varepsilon\}$, and box height $H$. (\textbf{c}) Plot showing the dependence of non-dimensional Brinkman medium resistance $\alpha\varepsilon$ on non-dimensional density $\rho\varepsilon^3$ of regularized Stokeslets.  Results are independent of the box height $H$. 
		}
		\label{bcflow}
	\end{center}
	\vspace{-20 pt}
\end{figure*}

\subsubsection{Analytical solution}
\vspace{-3mm}
Following the Brinkman equation, the resistance, $\alpha$, can determine the velocity field in the volume bounded by the plates as shown below. The fluid flow between the plates is governed by the Brinkman equations (Eq. \ref{brinkman}), and must also satisfy the velocity boundary conditions of $\mathbf{u} = U\hat{\mathbf{x}}$ at height $y = H$, and $\mathbf{u}=\mathbf{0}$ at $y = 0$. By symmetry there is only an $x$-component of the flow which obeys
\begin{equation}
-\frac{\partial p}{\partial x} = -\mu\left(\frac{\partial^2 u_x}{\partial x^2} + \frac{\partial^2 u_x}{\partial y^2} + \frac{\partial^2 u_x}{\partial z^2} - \alpha^2 u_x\right), 
\end{equation}    
where $u_x$ is the component of the velocity field in the $x$-direction. The left hand side of the equation is zero as there is no pressure gradient imposed in the $x$ direction. The first and third terms on the right hand side are zero by symmetry, since $\textbf{u}$ only depends on $y$. Solving for the velocity field from the balance of the remaining terms  subject to the boundary conditions ($u_x = U$ at $y = H$ and $u_x=0$ at $y=0$) gives the steady state flow profile
\begin{equation}
u_x(y) = \frac{U\sinh(\alpha y)}{\sinh(\alpha H)}.
\end{equation}
The resultant steady-state volume flow rate $Q$ through a central cross-section (as shown in Fig. \ref{bcflow}a) with width $L$ can be computed by integration of the flow field, ${Q = \int_{y=0}^{H} u_x(y)Ldy}$, to obtain 
\begin{equation}
Q = \frac{UL}{\alpha}\tanh\left(\frac{\alpha H}{2}\right). 
\label{qbcanal}
\end{equation}
\vspace{-3mm}
\subsubsection{Numerical setup}
\label{SecBEM}
\vspace{-3mm}
Next we model this flow using our proposed model with static regularized Stokeslets between the plates, and compute the volume flow rate through the central cross-section. Since the expression for volume flow rate (Eq. \ref{qbcanal}) depends on the resistance, $\alpha$, this will establish a relation between $\alpha$ and $\{\rho,\varepsilon\}$ of our proposed model. 
In order to be able to include the regularized Stokeslets numerically, we model the flow situation using a Boundary Element Method (BEM) \cite{Pozrikidis1992}.  In Appendix A, we present results of the validation of the BEM for a Newtonian fluid between two plates. 

First consider a box-shaped domain $D$ of width and length $L$  (see Fig. \ref{bcflow}a) between the two infinitely long parallel plates. The top and bottom surfaces of this domain move with velocity $U\hat{\textbf{x}} \text{ and } 0$ respectively. The right, left, front and back surfaces move with the local fluid velocity. According to the BEM, the Stokes flow at $\mathbf{x_0}$ inside a bounded domain, $D$, or its boundary, $\partial D$ (faces of the box in our case), can be expressed in terms of surface forces ($\mathbf{F}$) and velocities ($\mathbf{u}$) on the boundary via Boundary Integral Equations \cite{Pozrikidis1992}:
\vspace{-1mm}
\begin{eqnarray}
\frac{u_j(\textbf{x}_0)}{\beta} = \frac{1}{8\pi\mu}&&\int_{\partial D} S_{ij}(\textbf{x}-\textbf{x}_0)F_i(\textbf{x})dA \nonumber \\ -\frac{1}{8\pi}&&\int_{\partial D}u_i(\textbf{x})T_{ijk}(\textbf{x}-\textbf{x}_0)n_kdA, \label{BEMI} 
\end{eqnarray}
where 
\begin{equation}
\beta = \begin{cases}
1 & \;\;\;\;\; \mathrm{for}\; \mathbf{x}_0 \in D,\\
2 & \;\;\;\;\; \mathrm{for}\; \mathbf{x}_0 \in \partial D,
\end{cases}\label{bval} 
\end{equation}
$\mu$ is the fluid viscosity,  ${S_{ij}(\textbf{r}) =  \left(\delta_{ij}/{|\textbf{r}|} + {r_ir_j}/{|\textbf{r}|^3} \right)}$ is the (singular) Stokeslet,  ${T_{ijk}(\mathbf{r}) = {-6r_jr_jr_k}/{|\mathbf{r}|^5}}$ is the stresslet, and $n_k$ is the unit normal of the boundary pointing into the domain. We use indicial notation throughout this manuscript where $\{i,j,k\} = \{1,2,3\}$ run over cartesian directions, and repeated indices are implicitly summed over. We discretized the domain boundary ($\partial D$) into triangular elements \cite{Persson2006} and use Gaussian quadrature rules to numerically compute the surface integrals in the above equation. When there are $N_p$ regularized Stokeslets in the fluid domain, the fluid velocity at a point $\textbf{x}_0$ is then calculated as
\begin{eqnarray}
\frac{u_j(\textbf{x}_0)}{\beta} =&& \frac{1}{8\pi\mu}\sum_{b=1}^{N_b}\sum_{g=1}^{N_g}w^{(g)}S_{ij}(\textbf{x}^{(b,g)}-\textbf{x}_0)F_i(\textbf{x}^{(b,g)}) \Delta A^{(b)} \nonumber\\
&&-  \frac{1}{8\pi}\sum_{b=1}^{N_b}\sum_{g=1}^{N_g}w^{(g)} u_i(\textbf{x}^{(b,g)})T_{ijk}(\textbf{x}^{(b,g)}-\textbf{x}_0)n_k \Delta A^{(b)} \nonumber\\
&&+ \sum_{p=1}^{N_p} S_{ij}^\varepsilon(\textbf{x}^{(p)}-\textbf{x}_0)f^\varepsilon_i (\textbf{x}^{(p)}), \label{BEMON}
\end{eqnarray}
where $N_b$ is the number of boundary elements, $N_g$ is the number of Gaussian quadrature points with weights $w^{(g)}$, the value of $\beta$ depends on the position $\mathbf{x}_0$ according to the Eq. \ref{bval}, $\textbf{x}^{(b,g)}$ is the $g^{th}$ quadrature point of $b^{th}$ triangulated boundary element, $\Delta A^{(b)}$ is the area of $b^{th}$ boundary element, $N_p$ is the number of regularized Stokeslets representing the proposed porous medium model, $\textbf{x}^{(p)}$ is the position of the $p^{th}$ regularized Stokeslet, and $\mathbf{f}^\varepsilon(\textbf{x}^{(p)} )$ is the force acting on the $p^{th}$ regularized Stokeslet. We used 33 Gaussian quadrature points, and assumed constant force and velocity on each boundary element. 

Applying Eq. \ref{BEMON} with $\mathbf{x}_0$ at the centers of each boundary element yields $3 N_b$ equations in terms of the $3 N_b$ components of boundary forces, $3 N_b$ components of boundary velocities, and $3 N_p$ regularized Stokeslet forces. Applying Eq. \ref{BEMON} at the locations of the regularized Stokeslets ($\mathbf{x}_0 = \mathbf{x}^{(p)}$) yields $3 N_p$ additional equations in terms of the same variables, once the velocities at the regularized Stokeslet locations is set to zero to satisfy the static condition. 
Physically, as the regularized Stokeslets are fixed in place, the forces required to keep them stationary produces the effect of the porous medium on the flow.	
The solution to the above system of equations then requires $3 N_b$ additional conditions, which can be obtained by specifying  three of the force and velocity components $\{F_x, F_y, F_z, u_x, u_y, u_z\}$ for each boundary element, as described in what follows.

First, from the problem definition, the top plate moves to the right with velocity $U$ and the bottom plate is stationary. Thus at the top face of the box 
\begin{eqnarray}
u_x &=& U, \nonumber \\
u_y &=& 0, \nonumber\\
u_z &=& 0,\label{top}
\end{eqnarray}
and at the bottom face of the box
\begin{eqnarray}
u_x &=& 0, \nonumber\\
u_y &=& 0,\nonumber \\
u_z &=& 0. \label{bottom}
\end{eqnarray}
Furthermore, as there is no imposed pressure gradient, we choose the value of pressure such that $F_x = 0$ on the right and left faces. 
Strictly speaking, the force $F_x$ which is zero is the volume-averaged force (similar to the volume-averaged velocities and pressures in the Brinkman equation), not the microscopic force affected by the randomly placed regularized Stokeslets.  In the below we also prescribe the rest of the boundary conditions in terms of volume-averaged, macroscopic quantities.  This is justified post hoc by the independence of the results from the surface placement, i.e., the geometry of the box.

The symmetries of the problem determine enough of the remaining boundary forces and velocities to solve the problem. The box-shaped domain is an arbitrary portion of width and length $L$ between the two infinite plates. Thus the solution in the box has translational symmetry in the $x$ and $z$ directions, and is also symmetric when reflected about the $xy$-plane.
Due to translational symmetry the velocity field only depends on $y$, so the incompressibility condition ($\nabla\cdot \mathbf{u} = 0$) and the boundary conditions at the top and bottom plates imply that $u_y$ is zero all along the right, left, front, and back faces.
Together, reflection and translational symmetry imply that $u_z$ is zero on the right and left faces of the box: for a $u_z$ at some point on the right or left face, reflection about the $xy$-plane implies that the velocity at its reflected image location is $-u_z$, but the velocity at the image location is also $u_z$ by translation symmetry, hence must be zero. The same argument applies to $F_z$ on the right and left faces, and to $u_z$ on the front and back faces. 
Translation symmetry in the $z$-direction implies that the stress tensor is the same on front and back faces. Since the direction of normal changes sign for these faces, the direction of traction forces on the faces also changes sign. However, reflection symmetry about the $xy$-plane implies that $F_x$ and $F_y$ are the same on the front and back faces, so they must be zero.  Note that on the front and back faces symmetry does not prohibit a non-zero $F_z$, which corresponds to a normal stress difference.

Collecting these together, we know that on the right and left faces, $F_x=0$, $u_y=0$, $u_z=0$, and $F_z=0$.  On the front and back faces, we know that $F_x=0$, $F_y=0$, $u_y=0$, and $u_z=0$.  To solve the linear system we only need to specify six of these conditions in addition to Eqs. \ref{top}-\ref{bottom}. In the following we chose $F_x=0$, $u_y=0$, and $u_z=0$ on the right and left faces, and $F_x=0$, $F_y=0$, and $u_z=0$ on the front and back faces, and solve for all other variables.  In the Appendix, we validate this BEM method and choice of conditions for the case in which there is Newtonian fluid, not porous medium, between the plates.

\begin{figure*}[t]
	\begin{center}
		\includegraphics[width= 17cm]{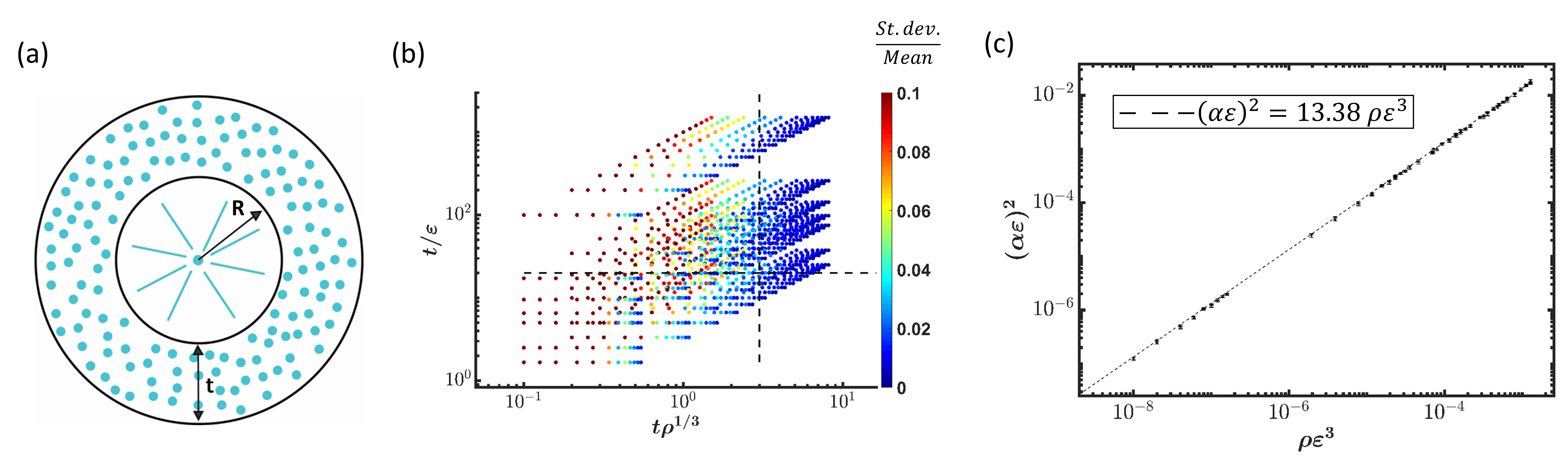}
		\vspace{-12 pt}
		\caption{(\textbf{a}) A point source with constant volume outflow, $Q$, is surrounded by a shell of porous medium, generating a pressure inside the shell. (\textbf{b}) Plot showing the ratio of standard deviation to the mean for estimates of $\alpha$  for a given combination of $\{\rho,\varepsilon\}$, and shell thickness $t$.  (\textbf{c}) Plot showing the dependence of non-dimensional Brinkman medium resistance $\alpha\varepsilon$ on the non-dimensional density $\rho\varepsilon^3$ of regularized Stokeslets. The results are independent of the shell geometry. 
		}
		\label{sourceflow}
	\end{center}
	\vspace{-20 pt}
\end{figure*}

The effective volume flow rate was computed by triangulating the central cross-section and computing area integral of velocity using one Gaussian quadrature point on each triangle (Eq. \ref{BEMON}) with $\beta = 1$, as the evaluation point is interior to the domain. The velocity field at these locations is the summation of velocity fields due to the boundary elements, and the forces at regularized Stokeslets representing porous medium. The net volume flow rate thus computed was compared to the analytical expression in Eq. \ref{qbcanal} to obtain a value of the resistance, $\alpha$, of the porous medium. Many such numerical experiments were conducted using different values of number densities ($\rho$), blob parameters ($\varepsilon$), and box heights ($H$) in the range $(64L^{-3},800L^{-3}) ,\,(2\times10^{-4}L,1.5\times10^{-2}L)\, \text{and } (0.1L,0.25L)$, respectively, to obtain an estimate of $\alpha$ for each case. 

In order for a random distribution of regularized Stokeslets to adequately represent a porous medium, we expect that the distribution must appear homogeneous on the scale of the porous medium geometry.  To investigate when this holds true, we plot in Fig. \ref{bcflow}b the coefficient of variation (ratio of standard deviation to the mean) obtained from the estimates of $\alpha$ as a function of the ratio of box height $H$ to blob size $\varepsilon$, and the ratio of box height $H$ to mean Stokeslet separation $\rho^{-1/3}$.
Each data point in Fig. \ref{bcflow}b is obtained from five different random distributions of regularized Stokeslets with a fixed density $\rho$ and blob size $\varepsilon$.
The coefficient of variation should be small if the distribution is homogeneous on the scale of the porous medium.
The plot shows that the coefficient of variation is small ($ <  0.05$) as long as $H/\varepsilon > 40$ and $H\rho^{1/3} > 2$ (dashed lines in Fig. \ref{bcflow}b), \textit{i.e.}, as long as the blob size and Stokeslet spacing is small enough compared to the geometry dimension. 
Note that for a fixed large epsilon ($H/\varepsilon < 40$), the coefficient of variation has nonmonotonic dependence on density, becoming large in the large $\varepsilon$, large $\rho$ regime.  In this regime, upon closer investigation we found that if we manually remove all regularized Stokeslets located within $\varepsilon$ of the boundary, the coefficient of variation becomes small.  Such a procedure changes the geometry; however, this behavior indicates that the large variations in this regime arise when regularized Stokeslets are randomly placed within a blob size of the boundary elements of the domain, which can lead to spatial variations in the velocities and tractions at the box surface, contrary to our prescription of boundary conditions as macroscopic quantities.

A plot of $\alpha$ against $\rho$, nondimensionalized by the length scale $\varepsilon$, is shown in Fig. \ref{bcflow}c.
In this plot, we only include data points corresponding to $H/\varepsilon > 40$, $H\rho^{1/3}>2$, i.e., with coefficient of variation $<0.05$.
 The resistance of a porous medium, $\alpha$, increases with increasing density and blob size of regularized Stokeslets. Each data point in Fig. \ref{bcflow}c is obtained by averaging $\alpha$ values computed from five different random distributions of regularized Stokeslets with a fixed density $\rho$ and blob size $\varepsilon$.  The error bars correspond to the standard deviation of those five values. 
The curves obtained for different values of $H/\varepsilon$ collapse to a single curve, indicating that the box geometry $H$ does not influence the value of $\alpha$ when the box dimensions are not comparable to the length scales involved in the regularized Stokeslets distribution.  This is a useful result, as we can estimate the value of $\alpha$ (resistance in Brinkman description) from the random distribution density ($\rho$) and blob size ($\varepsilon$) of regularized Stokslets in 3D. It can be seen that $\alpha\epsilon$ scales according to $(\alpha\epsilon)^2 \sim \rho\varepsilon^3$ by the linear fit to the data (obtained while forcing a zero intercept) in Fig. \ref{bcflow}c.

\subsection{Source flow}\label{numexp2}
We carried out a second numerical experiment to further corroborate the results obtained in Sec. \ref{numexp1}. Consider a source forcing a spherically symmetric flow through a shell of porous medium (Fig. \ref{sourceflow}a). A point source with constant volume outflow, $Q$, viscosity, $\mu$, is placed at the center of a porous shell of inner radius, $R$, and thickness, $t$. Due to the porous medium's resistance, constant pressure is generated inside the shell after a steady flow is established. 
\subsubsection{Analytical solution} 
The resistance, $\alpha$, of the Brinkman medium determines the pressure developed inside the shell as shown below. We solve Eq. \ref{brinkman} in the spherical coordinate system, where only the radial component is non-zero, 
\begin{equation}
-\frac{\partial p}{\partial r} = -\mu\left(\frac{1}{r^2}\frac{\partial}{\partial r}\left(r^2\frac{\partial u_r}{\partial r}\right)-\frac{2u_r}{r^2}\right)+\mu\alpha^2u_r,\label{bradial}
\end{equation}  
where $u_r$ is the radial component of the velocity $\mathbf{u}$ and $r$ is the radial distance. The solution is $u_r = Q/(4\pi r^2)$, which is same as in a Newtonian fluid due to incompressibility. Substituting the value of $u_r$ into Eq. \ref{bradial} and integrating from $R \text{ to }R+t$ gives an expression for the pressure generated across the thickness of the shell, $\Delta p$.  This expression can be used to compute the value of $\alpha$: 
\begin{equation}\label{kappaexpr}
\alpha^2 = \frac{4\pi}{\mu}\frac{\Delta p}{Q} R\left(1 + \frac{R}{t}\right).
\end{equation}

\subsubsection{Numerical setup}
Numerically, we represent the Brinkman medium by scattering static regularized Stokeslets in the thickness of the spherical shell. The constant volume outflow from the point source located at the center of the shell pushes these regularized Stokeslets radially outward with local fluid velocity, and consequently an opposing force is required to keep each of these regularized Stokeslets static. The flow field, $\textbf{u}$, at $\textbf{x}_0$ is
\begin{equation}
u_j(\textbf{x}_0) = \frac{Q}{4\pi |\textbf{x}_0|^3}{x_{0}}_j + \sum_{p=1}^{N_p} S_{ij}^\varepsilon(\textbf{x}^{(p)}-\textbf{x}_0)f^\varepsilon_i(\textbf{x}^{(p)})
\end{equation}
where $N_p$ is the number of regularized Stokeslets, $\textbf{x}_p$ is the location of the $p^{th}$ regularized Stokeslet, and $\textbf{f}^{\varepsilon}(\textbf{x}^{(p)})$ is the force acting on the $p^{th}$ regularized Stokeslet. Forces acting on each regularized Stokeslet can be computed by solving the system of equations obtained by applying the above equation with $\textbf{x}_0$ at the center of each regularized Stokeslet and imposing zero velocity on them [$\mathbf{u}(\mathbf{x}^{(p)}) = 0$]. These forces develop pressure, ${P}$, at the center of the shell according to: \cite{Cortez2002}
\begin{equation}\label{pressure}
P = \sum_{p = 1}^{N_p} \frac{2|\textbf{x}^{(p)}|^2+5\varepsilon^2}{\left(|\textbf{x}^{(p)}|^2+\varepsilon^2\right)^{5/2}}x^{(p)}_if^{\varepsilon}_i (\textbf{x}^{(p)} ).
\end{equation}
The pressure in the far field outside the spherical shell is zero. The numerical value of pressure difference across the porous shell thus obtained was compared to the analytical expression in Eq. \ref{kappaexpr}.

In Fig. \ref{sourceflow}b, we plot the coefficient of variation (ratio of standard deviation to the mean) obtained from the estimates of $\alpha$ as a function of the ratio shell thickness $t$ to blob size $\varepsilon$, and mean Stokeslet separation $\rho^{-1/3}$ for different values of $\rho, \, \varepsilon \, , \text{and } t$ in the range $(0.125 R^{-3}, 160R^{-3}), \, (10^{-3}R, 0.12R), \text{ and } (0.1R, 3R)$, respectively. 
Each data point is obtained from five different random distributions of regularized Stokeslets with a fixed density $\rho$ and blob size $\varepsilon$.
The coefficient of variation should be small if the distribution is homogeneous on the scale of the porous medium.
It can be seen that the coefficient of variation is small ($< 0.05$) as long as $t/\varepsilon> 20$ and $t\rho^{1/3}> 3$ (dashed lines in Fig. \ref{sourceflow}b).
Again, this shows that a random distribution of regularized Stokeslets can adequately describe the shell geometry as long as the blob size and Stokeslet spacing is small enough compared to a typical lengthscale of the geometry; otherwise, the distribution of Stokeslets is not sufficiently uniform to represent a Brinkman medium.

Using $\varepsilon$ as the length scale, $\alpha$ and $\rho$ are non-dimensionalized and plotted in Fig. \ref{sourceflow}c.
In this plot, we only include data points corresponding to $t/\varepsilon > 20$, and $t\rho^{1/3}>3$, i.e., with coefficient of variation $<0.05$.
Each data point is obtained by averaging $\alpha$ values computed from five different random distributions of regularized Stokeslets with a fixed density $\rho$ and blob size $\varepsilon$.  The error bars correspond to the standard deviation of those five values.
As for the Couette flow, it is encouraging to observe that the thickness ($t$) does not influence the non-dimensional $\alpha$. In other words, the experimental geometry does not influence the value of $\alpha$, and we get an estimate of the resistance of porous medium from only the density and blob size of regularized Stokeslets arrangement. It can be seen from Fig. \ref{sourceflow}c that $\alpha\varepsilon$ scales according to $(\alpha\varepsilon)^2 \sim \rho\varepsilon$ by the linear fit to the data (forcing a zero intercept).

\section{Discussion and conclusions}
\begin{figure}
	\includegraphics[width = 7cm]{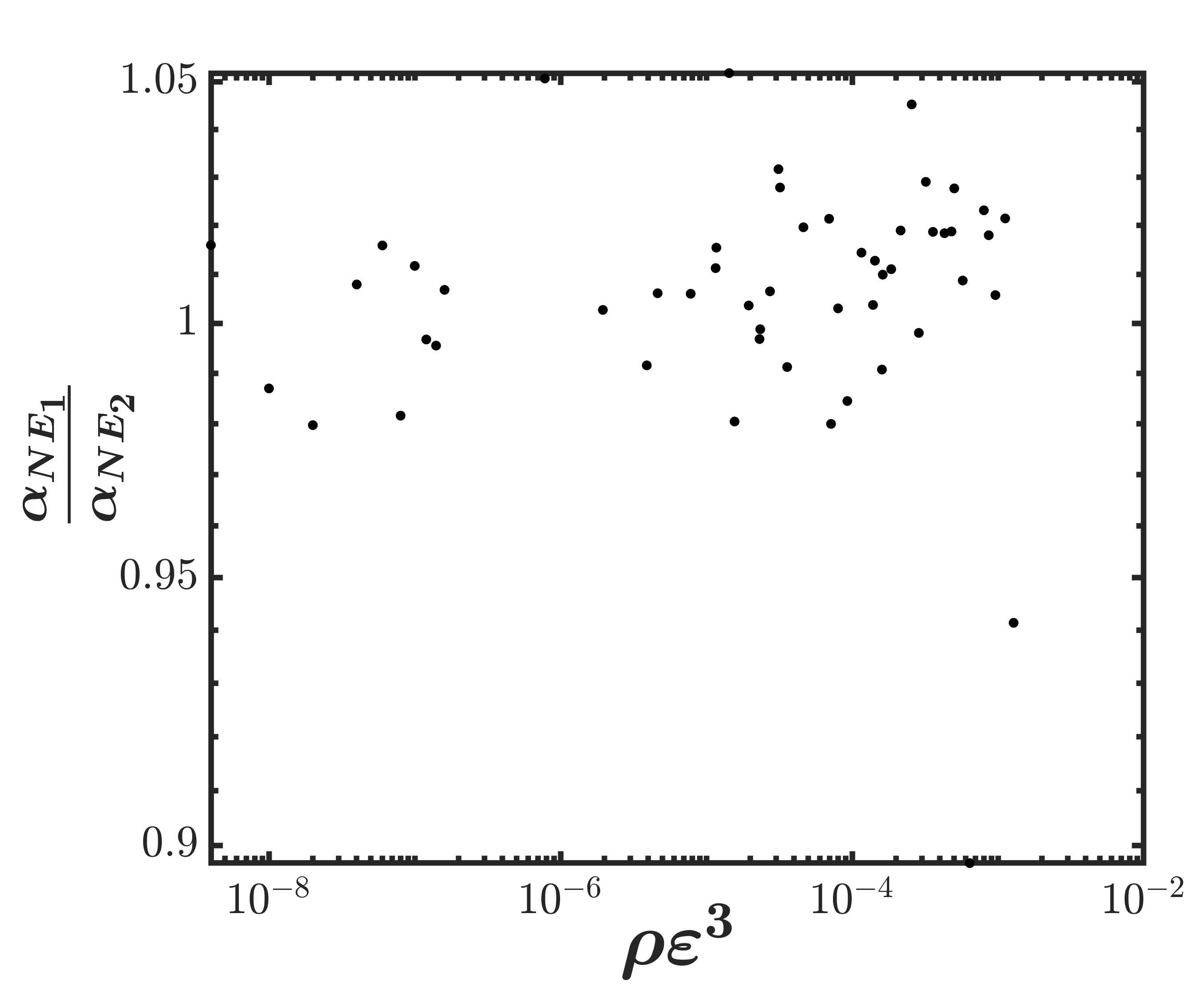}
	\caption{Plot showing the ratio of the resistance estimate $\alpha_{NE_1}$ from the first numerical experiment to the resistance estimate $\alpha_{NE_2}$ from the second numerical experiment, for each pair of density $\rho$ and blob size $\varepsilon$, plotted as a function of $\rho \varepsilon^3$.}\label{ratioplot}
\end{figure}

We have found that the square of Brinkman medium resistance $\alpha$, has a linear relation with the density $\rho$ and blob size $\varepsilon$ of regularized Stokeslets that represent the same porous medium. The results from the Couette flow and source flow experiments can be written as $\alpha = 3.72 \sqrt{\rho\varepsilon}$ and $\alpha = 3.66\sqrt{\rho\varepsilon}$, respectively. The difference in proportionality constants of these relations corresponds to $\sim2\%$ error in the estimates of $\alpha$ for a given $\rho$ and $\varepsilon$. The agreement between the two experiments can also be established by directly comparing the mean resistance estimates obtained from the two experiments at the same $\rho$ and $\epsilon$. 
We plot the ratio of estimates from the two numerical experiments against the parameter $\rho\varepsilon^3$ for each input pair $\{\rho,\varepsilon\}$ in Fig. \ref{ratioplot}. 
In this plot we only include the subset of datapoints from Figs. \ref{bcflow}c and \ref{sourceflow}c which have common values of $\rho$ and $\varepsilon$.
The ratios in Fig. \ref{ratioplot} are clustered around $1$, mostly within $~4\%$ error. This low difference in values of $\alpha$ from two independent numerical experiments corroborates the overall relation of $\alpha$ to $\sqrt{\rho\varepsilon}$. 

\begin{figure}
	\vspace{3mm}
	\includegraphics[width=7cm]{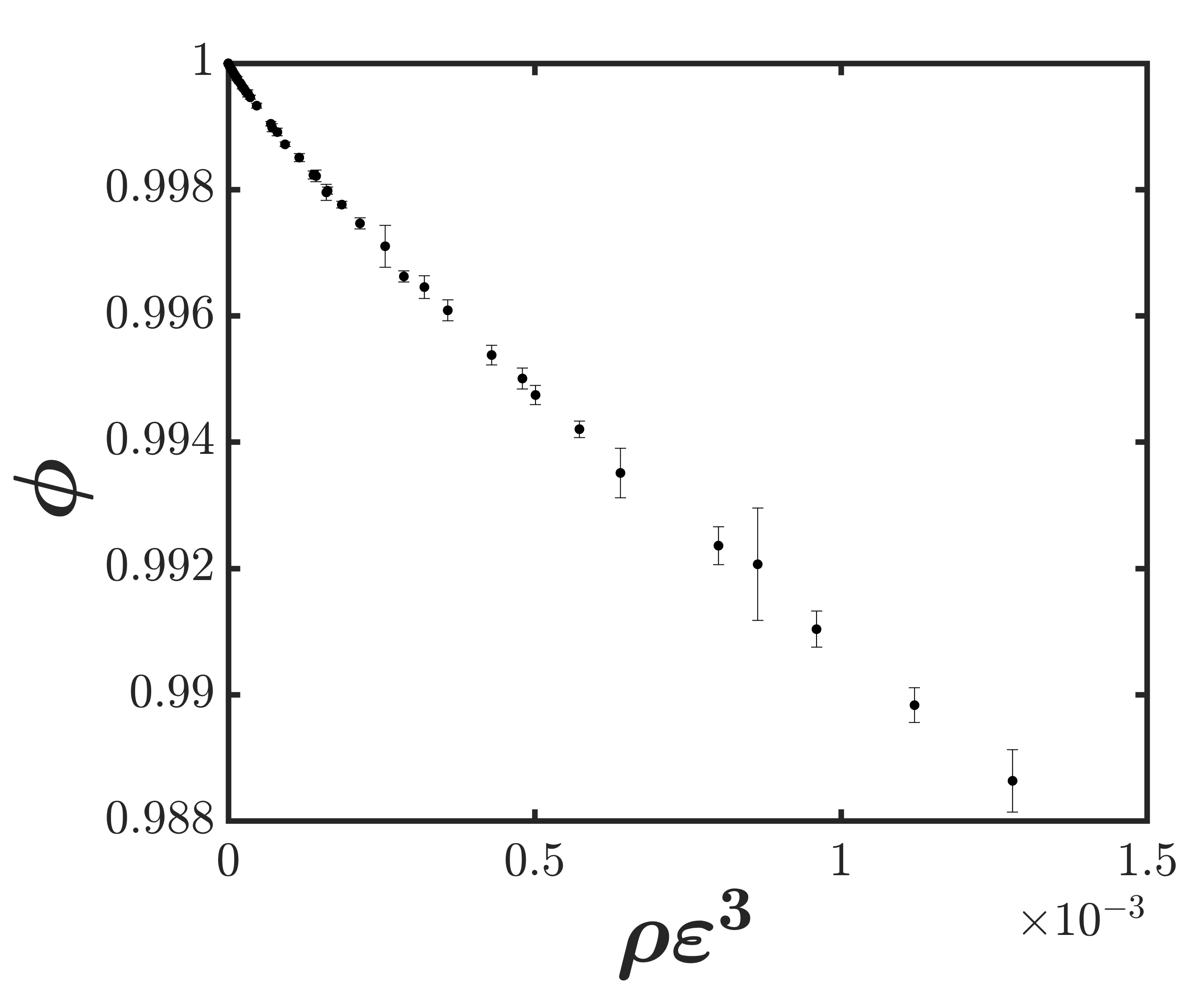}
	\caption{Volume fraction $\phi$ of fluid inside a Brinkman medium for a given regularized Stokeslet arrangment of density $\rho$ and blob size $\varepsilon$.}\label{phi}
\end{figure}

The volume fraction of the porous medium described by our random collection of regularized Stokeslets can be estimated from the results of 
Spielman and Goren \cite{Spielman1968}, who derived a formula for the volume fraction ($\phi$) of the fluid in terms of $\alpha$ for a random collection of fibers of radius $\varepsilon$ as 
\begin{equation}\label{phiformula}
\phi = \frac{\alpha\varepsilon K_0(\alpha\varepsilon) + 10K_1(\alpha\varepsilon)}{4\alpha\varepsilon K_0(\alpha\varepsilon)+10K_1(\alpha\varepsilon)},
\end{equation} 
where $K_0(.) \text{ and } K_1(.)$ are the zeroth- and first- order modified Bessel functions of the second kind. The same formula has been used \cite{Ho2019} to compute the volume fractions of biological fluids from values of $\alpha$. The result of applying this equation to the numerical experiments is shown in Fig. \ref{phi}. The error bars in the volume fraction plot are propagated from the error in the estimates of $\alpha$ from Sec. \ref{numexp1} and \ref{numexp2}. 

In this paper we have shown that the flows inside a porous medium which obey Brinkman equations can be described numerically by scattering many static regularized Stokeslets randomly in three dimensions. Using numerical experiments we have shown that we can estimate the resistance $\alpha$ of a corresponding Brinkman medium. The $\alpha$ of an equivalent Brinkman medium can be used to compute the permeability $\kappa \,(=\alpha^{-2})$ and the porosity $\phi$ (using Eq. \ref{phiformula}). Thus one can fully characterize the Brinkman medium from the values of $\rho$ and $\varepsilon$ of a regularized Stokeslet arrangement.
  
The main limitation on our method is that the lengthscales of the regularized Stokeslets distribution, $\{\rho^{-1/3},\varepsilon\}$ should be small compared to the typical lengthscales of the domain of porous media.  
For example, we found that sufficient homogeneity is provided when the blob size was less than a fortieth (twentieth) and the mean separation distance between the Stokeslets was less than half (a third) of the height of the box (shell thickness) in our first (second) numerical experiment. If there are other boundary elements or regularized Stokeslets describing domain geometries in the computation, one must also avoid situations where the regularized Stokeslets describing the porous media are likely to approach them closely, which occurs when the blob size and density are both large.  We observed such issues in the Couette flow but not in the source flow, which did not have any boundary elements describing its geometry.

Our model is applicable wherever a Brinkman medium is involved and is easy to implement as it just requires one to fill the volume occupied by the Brinkman medium with a random arrangement of static regularized Stokeslets of appropriate number density and blob size creating the desired resistance value $\alpha$. For instance, self-propelled swimming in a homogeneous Brinkman medium has been modeled using regularized Brinkmanlets for general \cite{Ho2019} and standard geometries \cite{Nganguia2018}; we can simulate this swimming using the method of regularized Stokeslets for the swimmer geometry \cite{Cortez2005} and scatter regularized Stokeslets at appropriate $\{\rho,\varepsilon\}$ around the swimmer to account for the effect of porous medium. 
Although the implementation of our method is simple if one is already using regularized Stokeslets, due to the large number of regularized Stokeslets it can quickly become computationally costly which might require the implementation of efficient solvers such as multipole methods \cite{Rostami2019, Yan2021}. Thus for homogeneous domains of porous media, the method of regularized Brinkmanlets may be more appropriate, but the real advantage of our model can be seen when there is a heterogeneous medium around a moving boundary. 
For example the actively self-generated confinement of bacteria swimming through gels has only been tackled analytically using extremely simplified approaches \cite{Mirbagheri2016, Reigh2017, Nganguia2020}, and cannot be treated with the method of regularized Brinkmanlets. Our method can be used to numerically treat this problem by filling the volume of the porous gel with static regularized Stokeslets, without having to account for the complicated boundary conditions near the medium interface arising due to the change in constitutive laws. This type of approach is also similar in spirit to the treatment of  viscoelastic fluids by a lattice arrangement of regularized Stokeslets interconnected by spring and dashpot elements \cite{Wrobel2014}.  

\begin{acknowledgements}
	S.K.K and H.C.F are supported by NIH Grant No.1R01GM131408-01. We also acknowledge the use of computing resources from the Center for High Performance Computing at the University of Utah.  
\end{acknowledgements}

\section*{Conflict of interest}

The authors declare that they have no conflict of interest.

\appendix
\section{Boundary conditions to model Newtonian shear flow inside a box}\label{newtonianval}
When there is no porous medium between the two infinitely long plates, the fluid flow there is governed by the Stokes equation. The velocity of the top plate sliding towards the right at velocity $U$ drives the fluid between the plates in a simple shear flow, $\mathbf{u}_{Newtonian} = U y \hat{\mathbf{x}}/H$. As in Sec. \ref{SecBEM} we model the fluid flow in a box of width and length $L$ via BEM where the top, bottom plate move with the prescribed velocity ($U \hat{\textbf{x}}$ and $0$ respectively). 

Similar to the main text, we discretize the boundary of the box into triangular elements and apply Eq. \ref{BEMON} with $\textbf{x}_0$ as the centers of each triangle, but with all regularized Stokeslet forces set to zero since there is no porous medium between the plates.  Then, we have a system of $3 N_b$ equations, in terms of $6 N_b$ components of traction force and velocity on each of the triangular elements.  
Thus for each side of the box, we can only specify three of $\{F_x,F_y,F_z,u_x,u_y,u_z\}$ and the other three are obtained from solving the system of equations.  All of these quantities can be calculated from the analytical solution, so below we test a large set of choices of which quantities are specified, and examine the result of the choice on the accuracy of the BEM for the flow field within the box.

On the top and bottom face of the box, $u_x$ comes from the problem definition, and motivated by the porous media problem, we consider $F_x$ to be an unknown output that must be calculated. Then there are 6 ways to pick the remaining two input quantities from $u_y$, $u_z$, $F_y$, and $F_z$. On the right and left faces of the box, we consider $u_x$ to be an unknown non-zero velocity field that must be calculated, and since $F_y$ is non-zero we do not prescribe its value. Then there are 4 ways to pick the three input quantities from $u_y$, $u_z$, $F_x$, and $F_z$.  On the front and back faces, we also consider $u_x$ to be an unknown velocity field to be calculated. Then there are 10 ways to pick the three input quantities from $u_y$, $u_z$, $F_x$, $F_y$, and $F_z$. In total, we check 240 ($6\times4\times10$) ways to prescribe the boundary conditions on the box in the BEM. The resulting system of equations can be arranged in the matrix form $AX=b$. Here $A$ is the coefficient matrix whose elements are the coefficients of unknown quantities in the BEM (Eq. \ref{BEMON}), $X$ is the column matrix with unknown quantities and $b$ is the column matrix computed by substituting the input quantities in the BEM (Eq. \ref{BEMON}). Then the unknown quantities are computed by solving the system of equations $AX = b$, and so we know the force and velocity of each triangular element representing the sides of the box. From these, we can compute the velocity field at different locations on the central cross section of the box using Eq. \ref{BEMON} with $\beta = 1$. 

Comparing the numerical solution to the analytical solution $\mathbf{u}_{Newtonian}$, we compute the norm error of the velocity field on the central cross section as $\sum_{n = 1}^{N_e}\left(|\mathbf{u}_{Newtonian}-\mathbf{u}(\mathbf{x}_{n})|A_n\right)^2$ and normalize with $(UL)^2$, where $N_e$ is the number of evaluation points, $\mathbf{x}_{n}$ is the location of $n^{th}$ evaluation point on the central cross-section, $\mathbf{u}$ is computed from Eq. \ref{BEMON}, $A_n$ is the area of $n^{th}$ element, and $|\mathbf{v}|$ is the magnitude of vector $\mathbf{v}$. For each of 240 possible ways to setup the boundary conditions, we computed the condition-number of the corresponding coefficient matrix $A$ and the resulting norm error and plot it in Fig. 5. The region on the bottom-left of the plot from condition number $5.4\times10^3$ to $3.1\times10^4$ corresponds to low condition numbers and minimum error in the velocity field. It can be seen that the boundary condition choices leading to low condition numbers give a low error in the velocity field. 
These accurate, low-condition number choices correspond to cases when for each side and each direction, either the velocity or force, but not both, are specified. For example, on the right and left sides, either $u_z$ or $F_z$ should be specified, in addition to $F_x$ ($u_x$ was considered to be an unknown) and $u_y$ (since $F_y$ was nonzero so we considered it to be an unknown). Note that the choice of conditions used for the porous media in the main text is one of these accurate and low condition number choices.

\begin{figure}
	\includegraphics[width=8cm]{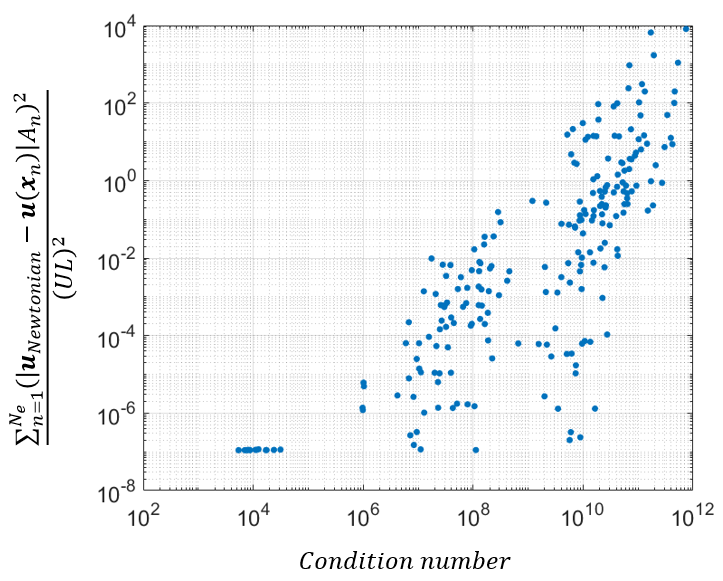}\label{l2error}
	\caption{Normalized error between the analytical velocity field and that computed from the boundary element formulation, for Newtonian Couette flow, plotted as a function of the condition number of coefficient matrix $A$ (see text) for different choices of boundary conditions on the box surface. We chose $U$ and $L$ as velocity and length scales, respectively, to normalize the error.}
\end{figure} 

\section*{Data Availability Statement}

The code required to generate the data reported to support the findings in this paper is published online along with this article. Additional information is available from the corresponding author upon reasonable request. 
\bibliography{library}

\end{document}